\documentclass[aps,pre,showpacs,amsfonts,amssymb,amsmath,floatfix,twocolumn,
floats,nobalancelastpage]{revtex4}

\usepackage{graphicx}

\begin{document}          

\title{Comment on ``Dynamic properties in a family of 
competitive growing models''} 

\author{A. Kolakowska}
\affiliation{Department of Physics and Space Sciences, 
Florida Institute of Technology, 150 W. University Blvd., 
Melbourne, FL 32901-6975}
\email{alice@kolakowska.us}

\author{M. A. Novotny}
\affiliation{Department of Physics and Astronomy, 
and HPC Center for Computational Sciences, 
Mississippi State University, 
P.O. Box 5167, Mississippi State, MS 39762-5167}

\date{\today}

\begin{abstract}
The article [Phys. Rev. E {\bf 73}, 031111 (2006)] by Horowitz 
and Albano reports on simulations of competitive surface-growth 
models RD+X that combine random deposition (RD) with another 
deposition X that occurs with probability $p$. The claim is made 
that at saturation the surface width $w(p)$ obeys a power-law 
scaling $w(p) \propto 1/p^{\delta}$, where $\delta$ is only 
either $\delta =1/2$ or $\delta=1$, 
which is illustrated by the models where X is ballistic deposition 
and where X is RD with surface relaxation. Another claim is 
that in the limit $p \to 0^+$, for any lattice size $L$, the
time evolution of $w(t)$ generally obeys the scaling
$w(p,t) \propto (L^{\alpha}/p^{\delta}) F(p^{2\delta}t/L^z)$,
where $F$ is Family-Vicsek universal scaling function.
We show that these claims are incorrect.
\end{abstract}


\maketitle

\begin{figure}[bp]
\includegraphics[width=8.0cm]{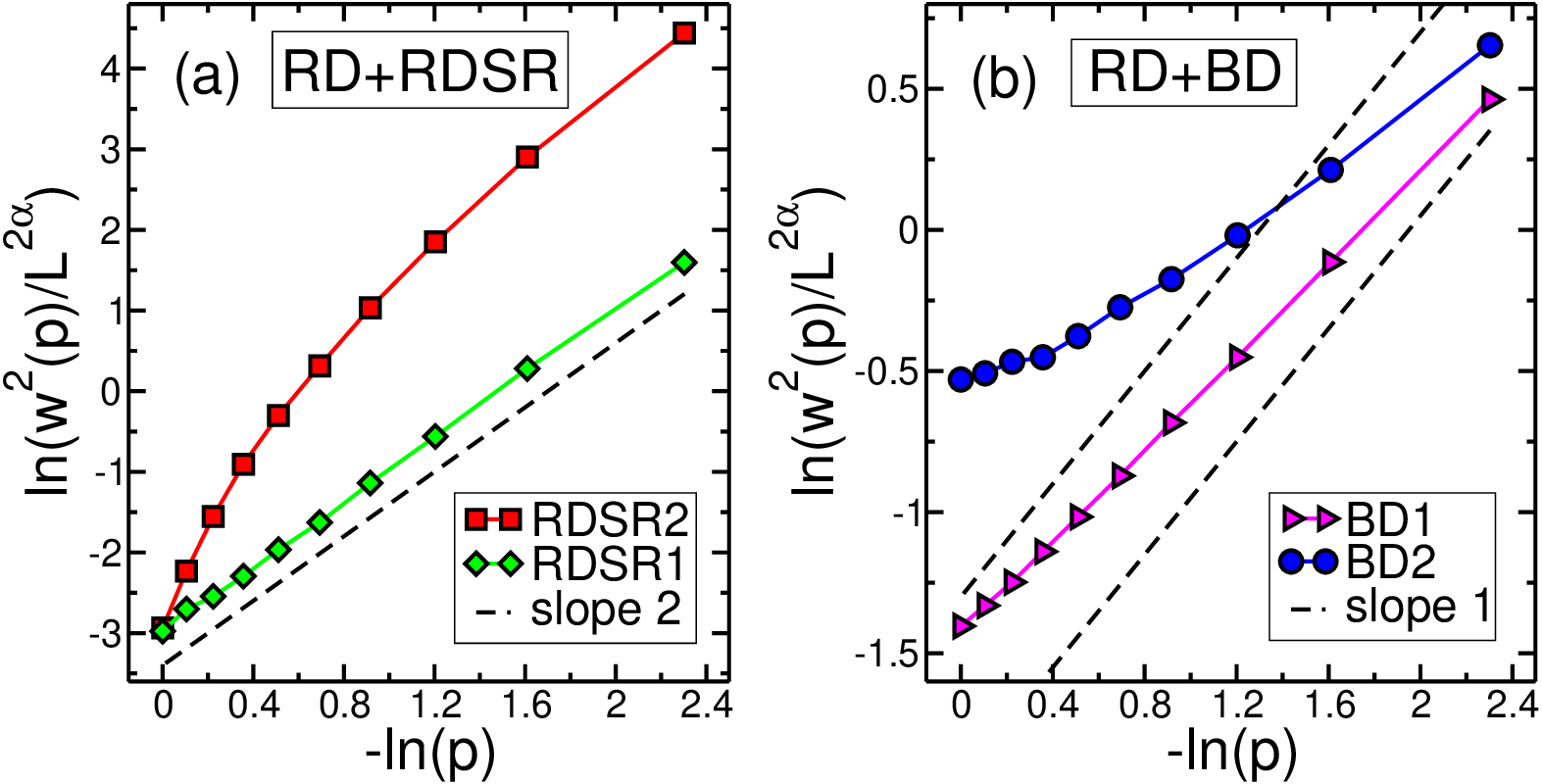}
\caption{\label{fig-1}
(color on line) 
Interface width at saturation in the RD+X model vs the selection probability 
$p$ of process X. 
(a) X is RDSR: the case when both RD and RDSR deposits are of unit  
height (diamonds, RDSR$1$; $L=500$); and, the case when RDSR deposits 
are of unit height and RD deposits are of twice that height 
(squares, RDSR$2$; $L=100$). 
(b) X is BD: the case of the NNN rule (circles, BD2); 
and, the case of the NN rule (triangles, BD$1$). In RD+BD simulations $L=500$.  
Solid line segments connecting data points (symbols) are guides for the eye. 
The dashed lines give reference slopes.
}
\end{figure}

\begin{figure}[t!]
\includegraphics[width=8.0cm]{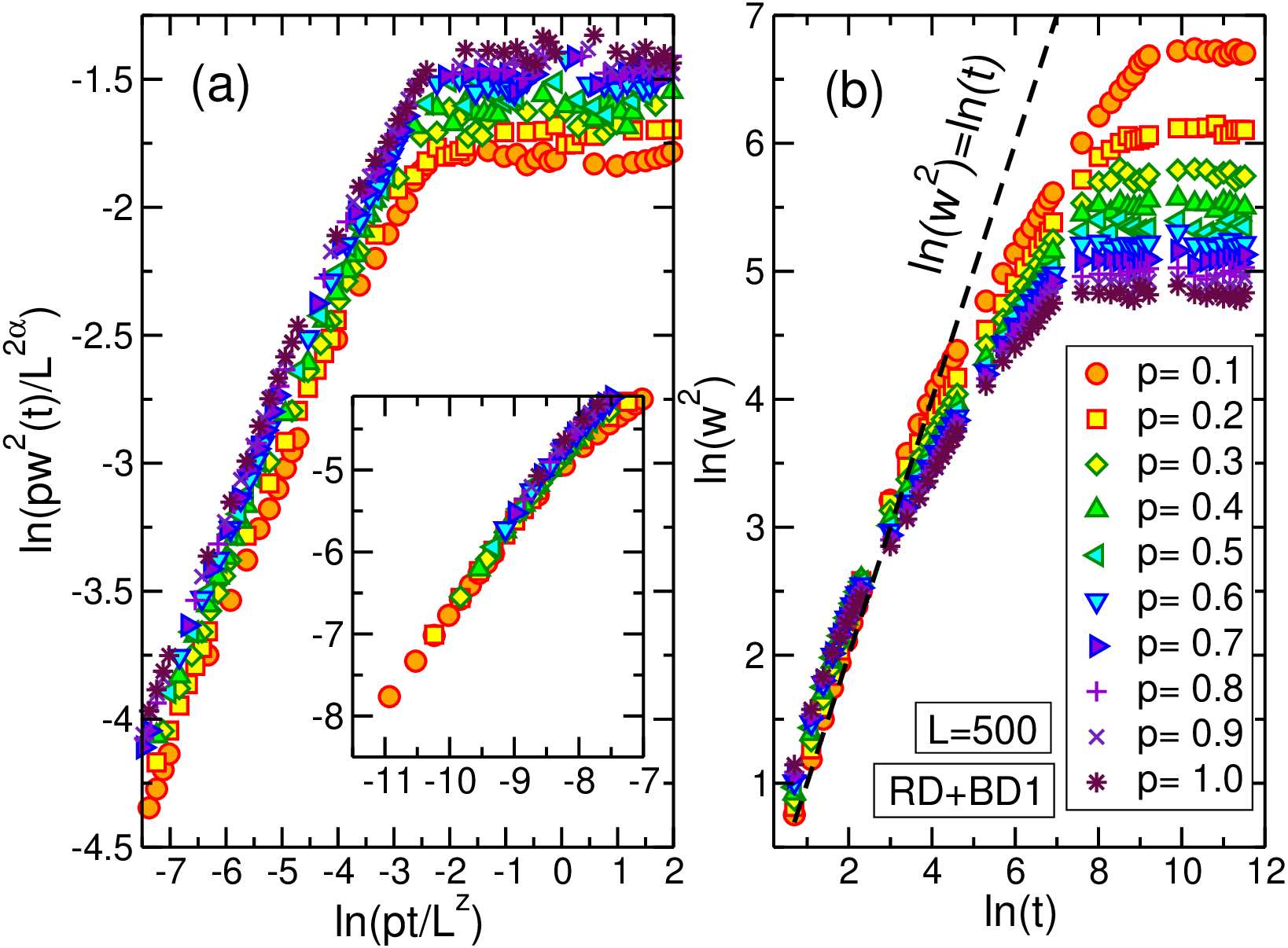}
\caption{\label{fig-2}
(color on line) 
Time-evolutions $w^2(p,t)$ in RD+BD$1$.
(a) Scaling in $p$ after Ref.\cite{HA06}. 
The insert shows the scaled initial transients.
(b) Evolution curves before scaling. The dashed line is
the RD evolution for $p=0$. In all models 
when the simulations start from a flat substrate $w(t)$ 
must pass an
initial transient before universal scaling can be measured.
The initial transients in part (b) follow RD universal evolution.
}
\end{figure}

\begin{figure}[t!]
\includegraphics[width=8.0cm]{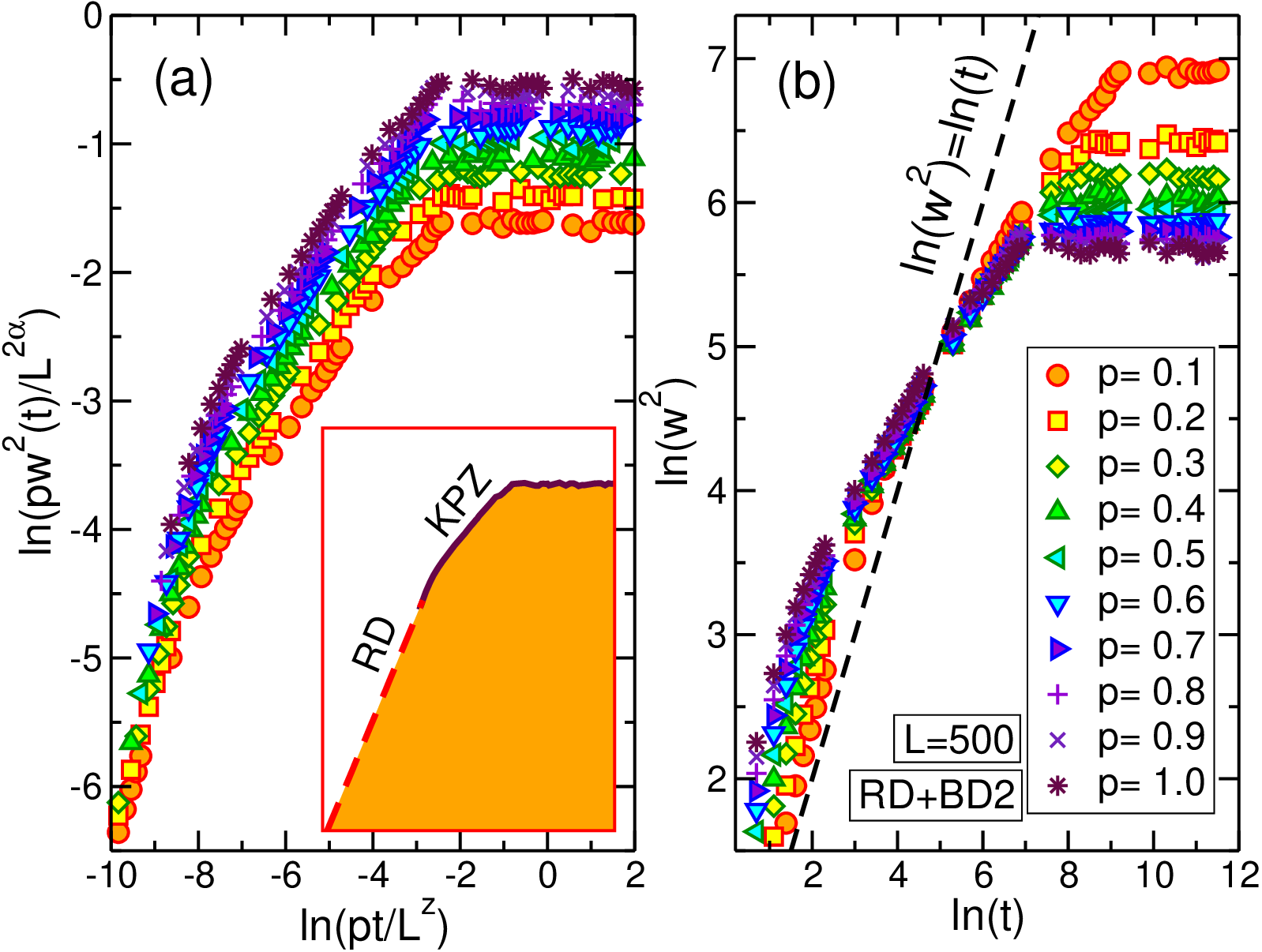}
\caption{\label{fig-3}
(color on line) 
Time-evolutions $w^2(p,t)$ in RD+BD$2$.
(a) Scaling in $p$ after Ref.\cite{HA06}. 
The outcome of this scaling is summarized in the insert.
(b) Data before scaling. The dashed line is the 
RD evolution for $p=0$.
}
\end{figure}

In Ref.\cite{HA06} the following scaling ansatz is proposed: 
\begin{equation}
\label{FV}
w^2(p,t) \propto \frac{L^{2\alpha}}{p^{2\delta}} 
F \left( p^{2\delta} \frac{t}{L^z} \right) \, ,
\end{equation}
where $w(p,t)$ are time evolutions of surface width in 
competitive growth models RD+X when a random deposition (RD) 
process is combined with process X, and $p \in (0;1]$ is the selection 
probability of process X. The function $F(\cdot)$ represents 
Family-Vicsek universal scaling. The anszatz (\ref{FV}) has been 
studied previously \cite{KNV04,KNV06,KN06} by examples where X 
represented either Kardar-Parisi-Zhang or 
Edwards-Wilkinson universal 
process. The new claim that is being made in Ref.\cite{HA06} is that 
a nonuniversal and \textit{model-dependent} exponent $\delta$ in 
Eq.(\ref{FV}) must be only of two values, either $\delta=1$ or 
$\delta=1/2$, for models studied in Ref.\cite{HA06}. 
To show that this 
claim is not correct we performed $(1+1)$ dimensional  
simulations of RD+X models when X is 
ballistic deposition (BD) and when X is random deposition with 
surface relaxation (RDSR), and performed scaling 
in accordance to Ref.\cite{HA06}. Our results are presented in 
figs.\ref{fig-1}-\ref{fig-3}.

Our data have been obtained on $L$ site lattices ($L$ is indicated 
in the figures) with periodic condition, starting from initially flat substrates, 
and averaged over $400$ to $600$ independent configurations. 
The time $t$ is measured in terms of the deposited monolayers. 
Simulations have been carried up to $t=10^7$, and the surface 
width at saturation has been averaged over the last 
$5000$ time steps. The data sets are for ten equally spaced selection 
probabilities $p$ from $p=0.1$ to $p=1$, where $p=0$ would 
be for RD process with no X present, and $p=1$ is for process X in 
the absence of RD. The data have been scaled in $L$ with the theoretical 
values of universal roughness exponent $\alpha$ and dynamic 
exponent $z$ of the universality class of process X. The RDSR 
algorithm used in our simulations is given in Ref.\cite{BS95} 
(Sec.5.1).  The BD algorithm used as X=BD1 is the nearest-neighbor (NN)  
sticking rule found in Ref.\cite{BS95} (Sec.2.2), and 
the BD algorithm used as X=BD2 is the next-nearest-neighbor (NNN) 
sticking rule found in Ref.\cite{BS95} (Sec.8.1).

\textbf{Saturation.} 
Saturation data (fig.\ref{fig-1}) show that in special cases 
an approximate power law $w(p) \propto 1/p^{\delta}$ may be observed. 
However, this is not a principle. Even if the data can be 
fit to the power law in $p$ only one of our examples shows a  
reasonable fit with $\delta \approx 1$ (seen in fig.\ref{fig-1}a). 
When X=BD$1$ the data in fig.\ref{fig-1}b show $\delta < 1/2$. 
The other two examples shown in fig.\ref{fig-1} defy a linear fit. 
In these cases there is no power law of the type claimed in 
Ref.\cite{HA06}. 
This absence of power-law scaling in $p$ is also evident in fig.4
of Ref.\cite{HA06}.

\textbf{The RD limit.} 
Another claim of Ref.\cite{HA06} is that Eq.(\ref{FV}) with 
the power-law prefactors $p^{\delta}$ 
(where $\delta=1$ or $1/2$) would prevail in the RD limit 
of $p \to 0^+$, and that such a scaling would be universal. 
We tested these claims in simulations of RD+BD models and 
found the evidence to the contrary (figs.\ref{fig-2}-\ref{fig-3}). 
In order to prove the absence of power-law scaling via 
Eq.(\ref{FV}) in the RD limit we present in 
figs.\ref{fig-2}b-\ref{fig-3}b the original $w^2(p,t)$ data 
before scaling. These original data show that parameter $p$, 
$p\in (0;1]$, assigns an order in the set of all curves 
$w^2(p,t)$ in such a way that $w^2(1,t)$ is the lowest 
lying curve, and at $p=0$ the initial transients become 
the RD universal evolution $w^2_{RD}(0,t) \propto t$.
The region between the boundaries $w^2(1,t)$ and $w^2_{RD}(0,t)$ 
is densely covered by the curves $w^2(p,t)$ because $p$ takes 
on continuous values. The pattern shown in 
figs.\ref{fig-2}b-\ref{fig-3}b for $p \in [0.1;1]$ extends down to 
values that are infinitesimally close to $p=0$, i.e., to 
the entire range of $p$. If the simulations are stopped 
at infinitesimally small $p'$ the width $w^2(p',t)$ is always the highest
lying curve in figs.\ref{fig-2}b-\ref{fig-3}b. In other words,
the smaller the $p'$ the higher the saturation value of $w^2(p',t)$.
But there is no bounding highest curve $w^2(p',t)$ in this set 
since the boundary $w^2(0,t)$ is the RD evolution. 
This order is reversed under the scaling of Eq.(\ref{FV}) when 
we set $\delta=1/2$, following Ref.\cite{HA06}. 
The outcome of this scaling is seen in 
figs.\ref{fig-2}a-\ref{fig-3}a: the boundary $w^2(1,t)$, i.e.,  
the lowest-lying curve in figs.\ref{fig-2}b-\ref{fig-3}b,  
is mapped onto the highest-lying curve in the image of this scaling 
seen in figs.\ref{fig-2}a-\ref{fig-3}a; and, a higher-lying 
curve $w^2(p,t)$ before scaling in 
figs.\ref{fig-2}b-\ref{fig-3}b is mapped onto a lower-lying curve 
after scaling in 
figs.\ref{fig-2}a-\ref{fig-3}a. 
In this scaling, the initial transients become ever longer 
as $p$ becomes ever smaller and closer to $p=0$, 
as seen in the insert in fig.\ref{fig-2}a. 
For any range of $p$, also in the limit $p \to 0^+$, 
the image of this scaling demonstrates no data collapse. 
This image is shown in the insert in fig.\ref{fig-3}a. 
Hence, for RD+BD models Eq.(\ref{FV}) with $\delta=1/2$ 
does not produce data collapse.

In some instances of model X, however, Eq.(\ref{FV}) 
can give an \textit{approximate} data collapse \cite{KNV06,KN06} 
but then $\delta$ is not restricted to the two values  
postulated in Ref.\cite{HA06}. For example, for the RD+BD1 model 
such scaling can be obtained with $\delta \approx 0.41$ \cite{KNV06} 
(note, $0.4<\delta<0.5$ is seen in fig.\ref{fig-1}b). But for the RD+BD2 
model there is no value of $\delta$ that would produce data 
collapse when nonuniversal prefactors in Eq.(\ref{FV}) 
are expressed as a power law $p^{\delta}$. We have 
demonstrated that such scaling does not generally exist 
and if occasionally it is observed it is a  
property of particular model.

In summary, the form of the nonuniversal prefactors as seen in 
universal Eq.(\ref{FV}) is a fit and is \textit{not} a principle.  
The exponent $\delta$ in Eq.(\ref{FV}) is model dependent, 
and the prefactor that enters may have other forms than $p^{\delta}$.

\end{document}